\documentclass[letterpaper, 10pt, conference]{ieeeconf}
\pdfoutput=1
\IEEEoverridecommandlockouts                             
\usepackage[utf8]{inputenc}
\usepackage{graphicx}
\usepackage{float}
\usepackage{subcaption} % For subfigures
\usepackage{amssymb}
\usepackage{amsfonts}
\usepackage{cite}

\usepackage{amsthm} % For theorems, assumptions, remarks, etc
\usepackage{flushend}

\usepackage{threeparttable}
\usepackage{tablefootnote}
\usepackage{multirow}
\usepackage{booktabs}

\makeatletter
\let\NAT@parse\undefined
\makeatother

\usepackage{hyperref}
\usepackage{xcolor}
\hypersetup{%
    colorlinks=true,
    linkcolor={blue!50!black},
    citecolor={blue!50!black},
    urlcolor={blue!50!black}
}

\usepackage{amsmath}

\newtheorem{remark}{Remark}

\usepackage[ruled, vlined, longend, linesnumbered, algo2e]{algorithm2e}
\SetKwInput{Input}{Input}
\SetKwInput{Hyperparam}{Hyper-params}
\SetKwInput{Return}{Approximate GNE of~\eqref{eq:GNEP}}

\usepackage{fancyhdr}
\newcommand{\R}{\mathbb{R}} % Real numbers
\newcommand{\N}{\mathbb{N}} % Natural numbers
\newcommand{\Ni}[2]{\N_{#1}^{#2}} % Interval of natural numbers
\newcommand{\cc}[1]{{\mathcal{#1}}} % Short for giving symbol the font for sets

\newcommand{\T}{^\top} % Transpose
\def\Sum#1#2{\sum\limits_{#1}^{#2}} % Sum with limits
\def\st{\rm{s.t.}} % For Subject To

\def\cX{\cc{X}} % Symbol for a state trajectory
\def\fracg#1#2{{\displaystyle{\frac{#1}{#2}}}} % Big \frac

\begin{document}
% Fakesection Frontmatter
\title{\LARGE \bf Learning generalized Nash equilibria from pairwise preferences}

\author{Pablo Krupa, Alberto Bemporad%
\thanks{This work was funded by the European Union (ERC Advanced Research Grant COMPACT, No. 101141351). Views and opinions expressed are however those of the authors only and do not necessarily reflect those of the European Union or the European Research Council. Neither the European Union nor the granting authority can be held responsible for them.
Corresponding author: Alberto Bemporad.
The authors are with the IMT School for Advanced Studies, Lucca, Italy. Email: {\tt\small \{pablo.krupa, alberto.bemporad\}@imtlucca.it}.}%
}

\maketitle
\thispagestyle{fancy} % For article submission
\pagestyle{fancy}

\begin{abstract}
Generalized Nash Equilibrium Problems (GNEPs) arise in many applications, including non-cooperative multi-agent control problems.
Although many methods exist for finding generalized Nash equilibria, most of them rely on assuming knowledge of the objective functions or being able to query the best responses of the agents.
We present a method for learning solutions of GNEPs only based on querying agents for their preference between two alternative decisions.
We use the collected preference data to learn a GNEP whose equilibrium approximates a GNE of the underlying (unknown) problem.
Preference queries are selected using an active-learning strategy that balances exploration of the decision space and exploitation of the learned GNEP.
We present numerical results on game-theoretic linear quadratic regulation problems, as well as on other literature GNEP examples, showing the effectiveness of the proposed method.
\end{abstract}

\begin{keywords}
    Game theory, Generalized Nash equilibrium, Preference-based learning, Machine learning, Active learning.
\end{keywords}

\section{Introduction} \label{sec:intro}

Generalized Nash Equilibrium Problems (GNEPs) arise in settings where multiple agents have conflicting, and possibly coupled, interests and constraints~\cite{Facchinei_AOR_2010}.
Some examples include economics~\cite{Dockner_GameEco_2000}, energy management~\cite{Hall_CDC_2022}, or control of multi-agent systems~\cite{Cappello_TCNS_2021}.
A Generalized Nash Equilibrium (GNE) is a solution of a GNEP in which no individual agent has an incentive to change its decision, given the decisions from other agents~\cite{Facchinei_AOR_2010}.
Finding a GNE is a challenging problem that has received a lot of attention from the research community, leading to the development of many methods with different convergence properties under assumptions of the objective functions and constraints of the GNEP.
When the objective functions of the agents are known, a GNE can be found using optimization-based methods such as operator splitting methods~\cite{Salehisadaghiani_CoRR_2017} or interior-point methods~\cite{Dreves_SJO_2011}.
Distributed methods have also been proposed~\cite{Salehisadaghiani_CoRR_2017, Tatarenko_TAC_2018, Carnevale_CDC_2024}, where each agent knows its own objective function, but full centralized knowledge is not required.

An alternative approach when full knowledge of the objective functions is not available is to learn a GNE using data obtained by querying either the objective functions or best responses of the agents.
Although literature on learning GNE from data is less extensive, several articles consider this setting.
In~\cite{Vorobeychik_AAMAS_2008}, the authors present a simulated annealing routine to learn an approximate Nash equilibrium using best response evaluations. 
In~\cite{Fabiani_TAC_2024} and~\cite{Franci_LCSS_2025} the authors propose Active Learning (AL)~\cite{prince2004} strategies for learning GNEs by querying best responses.
In~\cite{Picheny_JGO_2019}, equilibrium points of GNEPs with discrete search space are learned using a Bayesian optimization~\cite{Brochu2010} method that queries the objective functions using two proposed exploration-exploitation approaches.
An alternative Bayesian optimization approach to learn Nash equilibria, also based on querying objective function values, is presented in~\cite{Al-Dujaili_arXiv_2018}.
In~\cite{Nortmann_TAC_2024}, the authors propose an iterative data-driven approach to learn a GNE of the discrete-time game-theoretic Linear Quadratic Regulator (LQR) problem using random exploration of control actions.

A requirement of these data-based approaches is direct access to evaluations of the objective functions or best response solutions.
In contrast, preference-based learning~\cite{wirth_JMLR_2017} is a machine learning strategy that leverages information in the form of preferences, providing an alternative approach to learn a GNE when direct evaluations of the objective functions or best responses are unavailable.
Although preference-based learning has been mostly applied to train large language models~\cite{Ziegler2019FineTuningLM, Stiennon2020}, it has also been applied to control-related problems, such as tuning proportional-integral controllers~\cite{Coutinho2024_old} or model predictive controllers~\cite{Zhu_TCST_2022, Shao_LCSYS_24, Krupa_CDC_25}.

In this paper, we propose a preference-based learning approach to find a GNE that only relies on iteratively querying each agent for their preference between two different decisions, given the other agents' decisions.
We use the collected preference data to learn the objective functions of a GNEP whose equilibrium approximates a GNE of the underlying (unknown) GNEP.
To achieve this, we select preference queries using an AL strategy that balances exploration of the decision space and exploitation of the learned GNEP problem.
At each iteration of the proposed AL method, we query each agent for their preference, and then update the objective functions of the learned GNEP.
In contrast with previous works, the proposed approach does not require knowledge of the agents' objective functions, nor queries of the objective functions or best response of the agents.
Additionally, the method does not attempt to learn surrogate functions of the objective functions, which can be useful in settings where privacy is desired.
We present numerical results highlighting the effectiveness of the proposed method, both on GNEP problems taken from the literature, as well as on game-theoretic LQR problems.
An open source Python implementation of the proposed method is available at \url{https://github.com/pablokrupa/prefGNEP}.

The remainder of this paper is organized as follows.
Section~\ref{sec:learn} introduces the problem setting and shows how we apply preference-based learning to GNEPs.
The proposed AL preference-based method for finding a GNE is presented in Section~\ref{sec:AL}.
We show numerical results in Section~\ref{sec:results} and conclude the paper in Section~\ref{sec:conclusions}.

\noindent\textbf{Notation:}
Given $x, y \in \R^n$, $x \leq (\geq) \; y$ denotes componentwise inequalities.
The exponential function is denoted by $\exp(\cdot)$.
The set of natural numbers (including $0$) is denoted by $\N$.
For $i, j \in \N$ with $i \leq j$, $\Ni{i}{j} \doteq \{i, i+1, \dots, j\}$.

\section{Learning a GNE from preferences} \label{sec:learn}

Consider a GNEP with $n$ decision variables in which $N$ agents take decisions individually, each one responding with a decision vector
\begin{subequations} \label{eq:GNEP}
\begin{align}
    x_i^*(x_{-i}) \in \arg\min_{x_i \in \cc{X}_i} \; & J_i(x_i,x_{-i}) \\
\st \; & g(x)\leq 0, \\
   \; & h(x)=0,
\end{align}
\end{subequations}
where $x_i {\in} \R^{n_i}$ is the decision vector of agent $i$, $\sum_{i=1}^Nn_i{=}n$, $x_{-i}\in\R^{n-n_i}$ is the vector collecting all other agents' decisions $x_j$, $j \in \Ni{1}{N}$, $j\neq i$,
$x = (x_1, \ldots, x_N) \in \R^n$ is the stacked vector of all agents' decision variables\footnote{We may also write $x {=} (x_i, x_{-i})$ to highlight dependence on $x_i$ and~$x_{-i}$.},
functions $J_i \colon \R^{n_i} \times \R^{n - n_i} \to \R$ are the objective functions of each agent,
$g\colon\R^{n}\to\R^{n_g}$, $h\colon\R^{n}\to\R^{n_h}$ define global constraints,
and the compact sets $\cX_i \subset \R^{n_i}$ are local constraints.

A point $x^* = (x_1^*, x_2^*, \dots x_N^*)$ is a GNE of~\eqref{eq:GNEP} if, for each $i \in \Ni{1}{N}$, $J_i(x_i^*, x_{-i}^*) \leq J_i(x_i, x_{-i}^*), \; \forall x_i \in \cc{F}_i(x_{-i}^*)$, where
\begin{equation*}
\cc{F}_i(x_{-i}) {\doteq} \{ x_i \in \cc{X}_i \colon g((x_i, x_{-i})) \leq 0,\ h((x_i, x_{-i})) = 0 \}.
\end{equation*}

We assume that the objective functions $J_i(x_i,x_{-i})$ are unknown, and the constraint sets $\cX_i$ and functions $g$ and $h$ are known.
We take as a standing assumption that problem~\eqref{eq:GNEP} has at least one GNE.
Additionally, we consider that the best responses $x_i^*(x_{-i})$ cannot be measured directly.
However, we have access to measurements of preferences of the form
\begin{equation*}
    \pi_i(x_i^1,x_i^2;x_{-i}) = \left\{ \begin{array}{ll}
        1 & \text{if } J_i(x_i^1,x_{-i}) \leq  J_i(x_i^2,x_{-i}) \\
        0 & \text{otherwise,}   
    \end{array} \right.
\end{equation*}
which indicates the preference between two decision vectors $x_i^1$ and $x_i^2$ given the other agents' decisions $x_{-i}$.

Next, consider functions $\hat{J}_i \colon \R^{n_i} \times \R^{n - n_i} \to \R$ parameterized by $\theta_i \in \R^{n_{\theta_i}}$, and the resulting GNEP problem
\begin{subequations} \label{eq:GNEPi}
\begin{align}
    \hat{x}_i^*(\hat{x}_{-i}) \in \arg\min_{x_i \in \cc{X}_i} \; & \hat{J}_i(x_i,\hat{x}_{-i}; \theta_i) \\
\st \; & g(x)\leq 0, \\
   \; & h(x)=0.
\end{align}
\end{subequations}
Our objective is to learn parameters $\theta_i$ of functions $\hat{J}_i$ so that the GNE $\hat{x}^*$ of~\eqref{eq:GNEPi} approximates a GNE of~\eqref{eq:GNEP}.
Note that this objective is achieved if all $\hat{J}_i = J_i$, as~\eqref{eq:GNEP} and~\eqref{eq:GNEPi} would be the same.
That is, if we considered functions $\hat{J}_i$ as \emph{surrogate functions} that we train to match $J_i$ using data.
This approximation can be achieved using global optimization techniques, such as Bayesian optimization~\cite{Brochu2010}, if evaluations of the objective functions are available, see~\cite{Picheny_JGO_2019, Al-Dujaili_arXiv_2018}.
In this paper, however, we take an alternative approach that does not require direct queries of the objective functions nor best responses of the agents.
Instead, our approach is based on training functions $\hat{J}_i$ to obtain classifiers of the preferences between pairwise strategies as follows.\footnote{In the sequel we call functions $\hat{J}_i$ \emph{surrogate functions} for convenience, even though, again, the objective is not to train them to match functions $J_i$.}

Consider datasets $\cc{D}_{i} = \{ x_{i}^{j,1},x_{i}^{j,2},x_{-i}^{j}, \pi_{i}^{j} \}_{j = 0}^{M}$ for each agent $i \in \Ni{1}{N}$, where $M$ is the size of the dataset, $x_{i}^{j, 1}$ and $x_{i}^{j, 2}$ are the two options given to agent $i$ for the context $x_{-i}^j$, and $ \pi_{i}^{j} \doteq \pi_{i}(x_{i}^{j,1},x_{i}^{j,2};x_{-i}^{j})$ is the corresponding preference query.
We train $\theta_i$ so as to maximize the satisfaction of
\begin{equation} \label{eq:preference-surrogate}
    \pi_{i}(x_{i}^{j, 1}, x_{i}^{j, 2};x_{-i}^{j}) = 1 \iff \hat{J}_i^1 \leq \hat{J}_i^2, \; \forall j \in \Ni{1}{M},
\end{equation}
where $\hat{J}_i^1 \doteq \hat{J}_i(x_{i}^{j, 1},x_{-i}^{j}; \theta_i)$ and $\hat{J}_i^2 \doteq \hat J_i(x_{i}^{j, 2},x_{-i}^{j}; \theta_i)$.
We do this by posing a logistic regression classification problem.
Consider the sigmoid function
\begin{equation} \label{eq:prob}
    P_i(x_i^1, x_i^2, x_{-i}) = \frac{1}{1 + \exp\left(\fracg{\hat{J}_i^1 - \hat{J}_i^2}{d_i(x_i^1, x_i^2)} \right)},
\end{equation}
where $d_i\colon \R^{n_i} \times \R^{n_i} \to \R$ is a dissimilarity function that we add to improve classification accuracy when $x_i^1 \simeq x_i^2$ (see Remark~\ref{rem:dissimilarity}).
We take
\begin{equation} \label{eq:dissimilarity:log}
    d_i(x_i^1, x_i^2) = \log(\| x_i^1 - x_i^2\|_\infty + 1 + \epsilon_d),
\end{equation}
where $\epsilon_d > 0$ is some small number that is added to avoid division by $0$ in~\eqref{eq:prob}.
Using~\eqref{eq:prob}, we pose the learning problem
\begin{equation} \label{eq:learn:prob}
    \min\limits_{\theta_i \in \Theta_i} r_i(\theta_i) + \frac{1}{M} \Sum{j = 1}{M} \cc{L}(\pi_i^j, P_i(x_i^{j, 1}, x_i^{j,2}, x_{-i}^{j})),
\end{equation}
where $r_i\colon \R^{n_{\theta_i}} \to \R$ is a regularization term for $\theta_i$ (typically an $\ell_2$ or $\ell_1$ penalization, e.g., $r_i(\theta_i) = \rho_i \| \theta_i \|_2^2$ for $\rho_i > 0$),
$\cc{L}\colon \R \times \R \to \R$ is the cross-entropy loss 
\begin{equation*} \label{eq:learn:func:cross}
        \cc{L}(p, \hat{p}) = -p \log(\hat{p}) - (1 - p) \log(1 - \hat{p})
\end{equation*}
measuring the likelihood of the prediction $\hat{p} \in \R$ matching the target $p \in \{0, 1\}$,
and $\Theta_i \doteq \{\theta_i \colon \underline{\theta}_i \leq \theta_i \leq \overline{\theta_i}\}$ are non-empty (and possibly non-compact) box constraints on the parameters $\theta_i$.
Problem~\eqref{eq:learn:prob} can be solved using standard machine learning tools and solvers, e.g. L-BFGS-B~\cite{Byrd_LBFGS_JSC_1995}, Adam~\cite{Kingma_adam_2014}, or a combination of both~\cite{Bemporad_TAC_jax-sysid_25}.

We found that we can learn a GNE of~\eqref{eq:GNEP} by making $\hat{J}_i$ be a good local preference classifier around a GNE of~\eqref{eq:GNEP}.
Indeed, note that the learning problem~\eqref{eq:learn:prob} is a logistic classification problem in which $\hat{J}_i$ are trained to classify the preferences between the pairwise strategies in the datasets $\cc{D}_i$.
If datasets $\cc{D}_i$ contain pairs that are close to a GNE of~\eqref{eq:GNEP}, then we can expect the $\hat{J}_i$ obtained from~\eqref{eq:learn:prob} to be good local classifiers of the GNE.
The following section presents an AL loop whose objective is first to make $\hat{J}_i$ be a general approximation of $J_i$ by exploring the whole decision space, so that solutions of~\eqref{eq:GNEPi} resemble solutions of~\eqref{eq:GNEP}, and then to make $\hat{J}_i$ be a good local classifier of a GNE of~\eqref{eq:GNEP}, while simultaneously making solutions of~\eqref{eq:GNEPi} converge towards solutions of~\eqref{eq:GNEP}.

\begin{remark}[On the existence of a GNE for problem~\eqref{eq:GNEPi}] \label{rem:existance_GNE_surr}
A requirement of the proposed method is for problem~\eqref{eq:GNEPi} to possess a GNE solution.
We note that $g$, $h$ and $\cc{X}_i$ are assumed to be known.
Therefore, under certain constraint qualification conditions, functions $\hat{J}_i$ and the constraint sets $\Theta_i$ of parameters $\theta_i$ may be selected to ensure this.
We refer the reader to~\cite{Facchinei_AOR_2010} for conditions on the existence of a GNE.
\end{remark}

\begin{remark}[On the inclusion of the dissimilarity functions] \label{rem:dissimilarity}
One of the contributions of this paper is the inclusion of the dissimilarity functions $d_i$ in~\eqref{eq:prob}.
The reason for their inclusion is to improve the classification accuracy of the learned functions $\hat{J}_i$ between points $(x_i^1, x_{-i})$ and $(x_i^2, x_{-i})$ in which $\| x_i^1 - x_i^2\|$ is small, without affecting its accuracy when $\| x_i^1 - x_i^2\|$ is larger.
Note that, if $\| x_i^1 - x_i^2\| = 0$, then $P_i(x_i^1, x_i^2, x_{-i}) = 0.5$ for any function $\hat{J}_i$, as you always have $\hat{J}_i^1 = \hat{J}_i^2$.
Similarly, if $\| x_i^1 - x_i^2\|$ is very small, the minimization of the cross-entropy loss leads to a function $\hat{J}_i$ with a very large Lipschitz constant around $x_i^1$, so that small differences in the function argument push $P_i$ towards either $1$ or $0$ (note that $\pi_i$ is either $1$ or $0$).
We include the dissimilarity functions $d_i$ to alleviate this issue.
When $\| x_i^1 - x_i^2\|$ is small, we have that $d_i(x_i^1, x_i^2)$ is small, and thus we allow small differences in $\hat{J}_i$ to indicate strong preference between $x_i^1$ or $x_i^2$, according to the preference model~\eqref{eq:prob}.
On the other hand, as $\| x_i^1 - x_i^2\|$ increases, we require a larger difference between the values of $\hat{J}_i$.
Finally, we note that in this paper we take $d_i$ as~\eqref{eq:dissimilarity:log}, as we find that it provides the best results, although other choices of the dissimilarity function are possible, such as $d_i(x_i^1, x_i^2) = \| x_i^1 - x_i^2\|_2 + \epsilon_d$, or $d_i(x_i^1, x_i^2) = \sqrt{\| x_i^1 - x_i^2\|_2} + \epsilon_d$.
\end{remark}

{
\setlength{\algomargin}{1.0em}
\begin{algorithm2e}[h]
      \DontPrintSemicolon
      \caption{AL for GNE preference-based learning}
      \label{alg:AL}
      \Hyperparam{$\delta > 0$, $\sigma, \underline{\delta}, \underline{\sigma} \geq 0$, $k_{\rm max}, p_\delta, p_\sigma \geq 1$.}
      \Input{Initial datasets $\cc{D}_i^0$, functions $\hat{J}_i$, $d_i$, $z_i$ and $r_i$.}
      Learn initial $\theta^0_i$ by solving GNEP~\eqref{eq:learn:prob} using $\cc{D}_i^0$.\;
      \For{$k = 1, 2, \dots, k_{\rm max}$}{
          Select $\delta^k$ and $\sigma^k$ using~\eqref{eq:decay:delta},~\eqref{eq:decay:sigma} and Remark~\ref{rem:min_noise}.\;
          Obtain $(x_i^{k, 1}, x_{-i}^k)$, by solving the GNEP~\eqref{eq:GNEP:e-e}.\label{alg:AL:GNEP}\;
          Obtain $\hat{x}_i^{k, 2}$ by solving problems~\eqref{eq:e-e:best_response}. \label{alg:AL:BR}\;
          Obtain $x_i^{k, 2}$ using~\eqref{eq:sigma:noise}.\;
          Query preferences $\pi_i^k = \pi_i(x_i^{k, 1}, x_i^{k, 2}, x_{-i}^k)$.\;
          $\cc{D}_i^k \gets \cc{D}_i^{k-1} \cup \{(x_i^{k, 1}, x_i^{k, 2}, x_{-i}^k, \pi_i^k)\}$.\;
          Update $\theta^k_i$ by solving~\eqref{eq:learn:prob} using $\cc{D}_i^k$.\;
      }
      \Return{GNE of~\eqref{eq:GNEPi} using $\theta_i^{k_{\rm max}}$.}
  \end{algorithm2e}
}

\section{Active learning method for finding a GNE} \label{sec:AL}

Let $\cc{D}_{i}^0 = \{ x_{i}^{j,1},x_{i}^{j,2},x_{-i}^{j}, \pi_{i}^{j} \}_{j = 0}^{M_0}$ be the initial datasets, with $x_{i}^{j, 1}, x_{i}^{j, 2} \in \cc{X}_i$, and where $x^{j, \ell} = (x_i^{j, \ell}, x_{-i}^{j})$, $\ell \in \Ni{1}{2}$, satisfy global constraints, i.e., $g(x^{j, \ell}) \leq 0$ and $h(x^{j, \ell}) =~0$.
These initial datasets can be obtained by randomly sampling points satisfying the local and global constraints.

The proposed AL method is an iterative loop, where at each iteration $k$ we add a new sample to each dataset using the exploration-exploitation approach presented in the sequel.
This leads to datasets $\cc{D}_{i}^k = \{ x_{i}^{j,1},x_{i}^{j,2},x_{-i}^{j}, \pi_{i}^{j} \}_{j = 0}^{M_k}$, where $M_k = M_{k-1} + 1$.
At each iteration $k$, we solve the learning problem~\eqref{eq:learn:prob} using  datasets $\cc{D}_i^k$.
The AL loop presented in this section is summarized in Algorithm~\ref{alg:AL}, where ${k_{\rm max} \geq 1}$ is the user-selected number of AL iterations.
The output of Algorithm~\ref{alg:AL} is the proposed approximate GNE of~\eqref{eq:GNEP}.
We note that this learned approximate GNE always satisfies the local and global constraints, as it is a solution of~\eqref{eq:GNEPi}.

We consider pure exploration functions $z^k_i:\R^{n_i}\to\R$ for each agent $i$ that we want to maximize for promoting exploration of the decision space.
Examples of exploration functions are the IDW function~\cite{Bemporad_ML_21} based on the already collected data $\{x_{i}^{j, 1},x_{i}^{j, 2}\}_{j=1}^{M_k}$, or the space-filling function
\begin{equation} \label{eq:space_filling}
    z_i^k(x_i) = \min_{\ell \in \{1, 2\}, j \in \Ni{1}{M_k}} \| x_i - x_{i}^{j, \ell} \|.
\end{equation}
A simpler alternative is to consider the concave quadratic function
\begin{equation} \label{eq:simple_exploration}
    z_i^k(x_i) = -\frac{1}{2}\|x_i - \bar x^k_{i}\|_2^2
\end{equation}
where points $\bar{x}_i^k$ are either randomly sampled from $\cc{X}_i$ or $\bar x_i^k=\arg\max_{x_i \in \cc{X}_i} \bar z_i^k(x_i)$ and $\bar z_i^k$ is any of the aforementioned exploration functions.
At each iteration $k$, we generate a new query point $x^k = (x^k_1, \ldots, x^k_{N})$ by solving the GNEP%
\begin{subequations} \label{eq:GNEP:e-e}
\begin{align}
    x_i^{k}(x_{-i}^{k}) \in \arg\min_{x_i \in \cc{X}_i} \; & \hat J_i(x_i,x_{-i}^{k}; \theta_{i}^{k-1}) -\delta^k z_i^k(x_i)\\
\st \; & g(x)\leq 0, \\
\; & h(x)=0,
\end{align}
\end{subequations}
where $\delta^k>0$ is a weighting factor trading off between exploration and exploitation. 
A typical approach in AL methods is to start with a large exploration term that is reduced as $k$ increases, see, e.g.,~\cite{Zhu_TCST_2022}.
The idea is to promote exploration at the beginning of the AL loop, when the surrogate functions $\hat{J}_i$ are (presumably) poor candidates, and then rely more on exploitation of functions $\hat{J}_i$ as the iteration counter $k$ increases. 
In~\cite{Zhu_TCST_2022}, the authors propose a linear decay function for $\delta^k$.
However, we find that better results are typically obtained in our problem setting if we use an exponential decay function of the form
\begin{equation} \label{eq:decay:delta}
    \delta^k = \delta \left( 1 - \frac{k}{k_{\rm max}} \right)^{p_\delta},
\end{equation}
where $\delta > 0$ and $p_\delta \geq 1$ are user-selected hyperparameters.
If $p_\delta = 1$, we recover the linear decay rate used in~\cite{Zhu_TCST_2022}.
The datasets $\cc{D}_i^{k}$ are obtained by augmenting $\cc{D}_i^{k-1}$ with the new samples $(x_i^{k, 1}, x_i^{k, 2}, x_{-i}^{k}, \pi_i^{k})$, where $x_i^{k, 1} = x_i^{k}$ are the solutions obtained from solving~\eqref{eq:GNEP:e-e}, $\pi_i^{k} = \pi_i(x_i^{k, 1}, x_i^{k, 2}; x_{-i}^{k})$, and we take each $x_i^{k, 2}$ by first computing the best response for the $x^k_{-i}$ obtained from solving~\eqref{eq:GNEP:e-e}, according to the current surrogate function $\hat{J}_i$, i.e., by solving problems
\begin{subequations} \label{eq:e-e:best_response}
\begin{align}
    \hat{x}_i^{k, 2} \in \arg\min_{x_i \in \cc{X}_i} \; & \hat J_i(x_i,x_{-i}^{k}; \theta_{i}^{k-1}) \\
\st \; & g((x_i, x_{-i}^{k}))\leq 0, \\
\; & h((x_i, x_{-i}^{k}))=0,
\end{align}
\end{subequations}
and then taking
\begin{equation} \label{eq:sigma:noise}
    x_i^{k, 2} = \hat{x}_i^{k, 2} + \sigma^k w^k \| \hat{x}_i^{k, 2} \|_\infty,
\end{equation}
where each element of $w^k \in \R^{n_i}$ is sampled from a uniform distribution in the interval $[-0.5, 0.5]$, and
\begin{equation} \label{eq:decay:sigma}
    \sigma^k = \sigma \left( 1 - \frac{k}{k_{\rm max}} \right)^{p_\sigma}
\end{equation}
for some user-selected hyperparameters $\sigma \geq 0$ and $p_\sigma \geq~1$.
The proposed AL approach compares $x_i^{k, 1}$, which is the solution of a GNEP~\eqref{eq:GNEP:e-e} that combines the current surrogate $\hat{J}_i$ with an exploration term $z_i^k$, with $x_i^{k, 2}$, which is a noise-altered best response (without exploration term) for the solution obtained from~\eqref{eq:GNEP:e-e}.
If $\delta = \sigma = 0$, then $x_i^{k, 1} = x_i^{k, 2}$.

\begin{remark}[On adding noise to $x_i^{k, 2}$] \label{rem:noise}
We note that a more straightforward approach would be to take $x_i^{k, 2} = \hat{x}_i^{k, 2}$, i.e., as the best response for the solution $x_i^k$ of the GNEP~\eqref{eq:GNEP:e-e}.
Instead, we add some random noise to $x_i^{k, 2}$ in~\eqref{eq:sigma:noise}, normalized by the magnitude of $x_i^{k, 2}$.
The reason is that by adding this noise we obtain a vector $(x_i^{k, 2}, x_{-i}^k)$ that might violate constraints.
Although this might seem counter-intuitive, note that we approach the problem of learning a GNE of~\eqref{eq:GNEP} as a classification problem.
That is, we seek to learn functions $\hat{J}_i$ that locally classify preferences on $J_i$ around solutions of the GNEP~\eqref{eq:GNEPi}.
We therefore want functions $\hat{J}_i$ to classify correctly at the boundary of the constraints when~\eqref{eq:GNEP} has a GNE with active constraints.
Adding a (relatively small) random noise to $x_i^{k, 2}$ allows us to learn a GNE with active constraints, as it provides information about $J_i$ on points that are unfeasible but close to the boundary of the constraints.
The noise term $\sigma$ can be set to $0$ in problems where querying agents for non-admissible decisions is unreasonable.
\end{remark}

\begin{remark}[Minimum values of $\delta^k$ and $\sigma^k$] \label{rem:min_noise}
As discussed in Remark~\ref{rem:dissimilarity}, we include a dissimilarity term $d_i$ in~\eqref{eq:prob} to improve classification performance when $| \hat{J}_i^1 - \hat{J}_i^2|$ is small.
However, we can still run into numerical issues when $\hat{J}_i^1$ and $\hat{J}_i^2$ are too similar.
Note that, since $\delta^k \to 0$ and $\sigma^k \to 0$ as $k \to \infty$, we can expect $\|x_i^{k, 1} - x_i^{k, 2} \| \to 0$, and therefore that $| \hat{J}_i^1 - \hat{J}_i^2| \to 0$.
To avoid this, we set a user-defined minimum value $\underline{\delta} \geq 0$ for $\delta^k$, i.e., $\delta^k \gets \max(\delta^k, \underline{\delta})$.
We do the same for $\sigma^k$ with a user-defined~$\underline{\sigma} \geq 0$, taking $\sigma^k \gets \max(\sigma^k, \underline{\sigma})$.
\end{remark}

\section{Numerical results} \label{sec:results}

We present numerical results using the implementation of Algorithm~\ref{alg:AL} publicly available at \url{https://github.com/pablokrupa/prefGNEP}, which uses the \texttt{NashOpt} Python package~\cite{NashOpt} (version \texttt{1.1.0}) to solve the GNEP and best response problems in Steps~\ref{alg:AL:GNEP} and~\ref{alg:AL:BR} of Algorithm~\ref{alg:AL}.
The numerical results in this section can be reproduced by running the example scripts in the repository.
For the learning problem~\eqref{eq:learn:prob}, we take the regulation terms as $r(\theta_i) = 0.001 \| \theta_i \|_2^2$, and use $500$ iterations of Adam (with learning rate of $0.001$ and decay rates $\beta_1=0.9$ and $\beta_2 = 0.999$) followed by a maximum of $1000$ iterations of the L-BFGS-B solver from the \texttt{jaxopt} package (using a history size of $10$), initialized from the solution returned by Adam.
We use the dissimilarity functions~\eqref{eq:dissimilarity:log} and take the exploration functions $z_i^k$ as~\eqref{eq:simple_exploration}, with $\bar{x}_i^k$ randomly sampled from the local constraints $\cc{X}_i$.
We take the surrogate functions as 
\begin{equation} \label{eq:surr:small_quad}
    \hat{J}_i(x_i, x_{-i}; \theta_i) = \frac{1}{2}  x_i\T P_i x_i + q_i\T x_i + x_{-i}\T A_i x_i,
\end{equation}
where $\theta_i$ contains $q_i \in \R^{n_i}$, $A_i \in \R^{(n - n_i) \times n_i}$ and the non-zero elements of the Cholesky decomposition of the positive definite matrix $P_i \in \R^{n_i \times n_i}$.
Lower bounds $\underline{\theta}_i$ can be imposed on $\theta_i$ in~\eqref{eq:learn:prob} to ensure that $P_i$ is positive definite, see, e.g.,~\cite{Krupa_CDC_25}.
All the numerical results use $\sigma = 0.3$, $\underline{\delta} = 10^{-4}$, $\underline{\sigma} = 10^{-3}$,  $p_\delta = 5$ and $p_\sigma = 3$ as hyperparameters of Algorithm~\ref{alg:AL}, and use initial datasets of size $M_0 = 50$.

\subsection{Game-theoretic linear quadratic regulator} \label{sec:results:LQR}

Consider a discrete linear time-invariant system
\begin{equation} \label{eq:sys:NLQR}
    \xi(t + 1) = A \xi(t) + B u(t),
\end{equation}
where $\xi(t) \in \R^{n_\xi}$ and $u(t) \in \R^m$ are the state and control input at sample time $t$, $A \in \R^{n_\xi \times n_\xi}$ and $B \in \R^{n_\xi \times m}$.
Additionally, we have $N$ agents, each one controlling a portion of the control inputs, such that $u(t) = (u_1(t), u_2(t), \dots u_N(t))$, with $u_i(t) \in \R^{m_i}$ being the control action of agent $i \in \Ni{1}{N}$, and $\sum_{i = 1}^{N} m_i = m$.
The objective of each agent is to control~\eqref{eq:sys:NLQR} so as to minimize the Linear Quadratic Regulator (LQR) cost for matrices $Q_i \in \R^{n_\xi \times n_\xi}$ and $R_i \in \R^{m_i \times m_i}$, with $Q_i \succeq 0$ and $R_i \succ 0$, see, e.g.~\cite[\S II]{Nortmann_TAC_2024} or~\cite[\S 5.1]{NashOpt} for a more in-depth explanation of this problem setting.

It is well known that this game-theoretic LQR problem admits a solution of the form $u_i(t) = -K_i \xi(t)$,
where $K_i \in \R^{n_\xi \times m_i}$ is the state-feedback gain for agent $i$, see, e.g.,~\cite{Nortmann_TAC_2024}.
We compute approximate gains $K_i$ by solving the GNEP~\cite{NashOpt}
\begin{align} \label{eq:GNEP:LQR}
    K_i^*(K_{-i}) \in \arg \min_{K_i} \;& \Sum{j = 0}{T} \xi_j\T Q_i \xi_j + u_{i, j}\T R_i u_{i, j} \\
    \st \;& \xi_{j+1} = (A - B K_{-i}) \xi_j + B_i u_{i, j} \nonumber \\
    \;& u_{i, j} = -K_i \xi_j, \nonumber
\end{align}
where a more accurate solution of the infinite horizon LQR game is obtained as $T > 0 $ is increased.
Problem~\eqref{eq:GNEP:LQR} can be posed as~\eqref{eq:GNEP} by taking $J_i$ as the squared Frobenius norm of the deviation between $K_i$ and the best response $K_i^*(K_{-i})$ for agent $i$, i.e., as the best response deviation
\begin{equation} \label{eq:J:NLQR}
    J_i(K_i, K_{-i}) = \| K_i^*(K_{-i}) - K_i \|_F^2.
\end{equation}

We consider three examples, each with the number of agents $N$ and system dimensions $n_\xi = m$ shown in each row of Table~\ref{tab:LQR}.
We take $m_i = m / N$, $R_i = I_{m_i}$ and $Q_i$ as a matrix of zeros with ones in the diagonal elements corresponding to states $(i-1) m_i + 1$ to $i m_i$.
Matrices $A$ and $B$ of~\eqref{eq:sys:NLQR} are randomly generated so that $A$ is an unstable matrix with spectral radius equal to $1.1$.
Taking $T = 50$, we obtain a solution of problem~\eqref{eq:GNEP:LQR} using the \texttt{NashLQR} solver from the \texttt{NashOpt} Python package~\cite{NashOpt}, which considers~\eqref{eq:J:NLQR}.

For each example, we learn a GNE of the game-theoretic LQR problem using Algorithm~\ref{alg:AL} with $\delta = 5$, where $x_i$ and $x_{-i}$ are the vectorized form of $K_i$ and $K_{-i}$,
We denote by $\hat{K}^k_i$ the gain matrices learned by Algorithm~\ref{alg:AL} at iteration~$k$.

We note that we consider the objective functions~\eqref{eq:J:NLQR} to provide a numerical example in which the preference queries $\pi_i$ use the same objective function that is used in the \texttt{NashOpt} package, which we use as our back-end GNEP solver.
In a practical setting, the preference queries $\pi_i$ would be taken by asking each agent for their preference between two alternative controllers, e.g., by performing a closed-loop experiment with each controller.
Algorithm~\ref{alg:AL} only requires the preference information, i.e., the values of the underlying functions $J_i$ driving the preference decision are not required.

\begin{table}[t]
\centering
\vspace*{1em}
\begin{tabular}{l@{\ }l|cccc}
\multicolumn{2}{l}{} & \multicolumn{2}{c}{$k_{\rm max} = 100$} & \multicolumn{2}{c}{$k_{\rm max} = 200$} \\
\cmidrule(lr){3-4} \cmidrule(lr){5-6}
\multicolumn{2}{l}{Parameters} & RMSE & $\max_i J_i$ & RMSE & $\max_i J_i$ \\
\midrule
$n_\xi = m = 6,$  & $N=3$ & 0.00343 & 0.0970 & 0.00109 & 0.0202 \\
$n_\xi = m = 8,$ & $N=3$ & 0.00675 & 0.0596 & 0.00288 & 0.0144 \\
$n_\xi= m = 12,$ & $N=4$ & 0.01109 & 0.3009 & 0.01229 & 0.2152 \\ % 0.00900 & 0.1724 for seed 1 and k=200
\bottomrule
\end{tabular}
\caption{Results for game-theoretic LQRs}
\label{tab:LQR}
\end{table}

\begin{figure*}[t]
    \centering
    \begin{subfigure}[ht]{0.32\textwidth}
        \includegraphics[width=\linewidth]{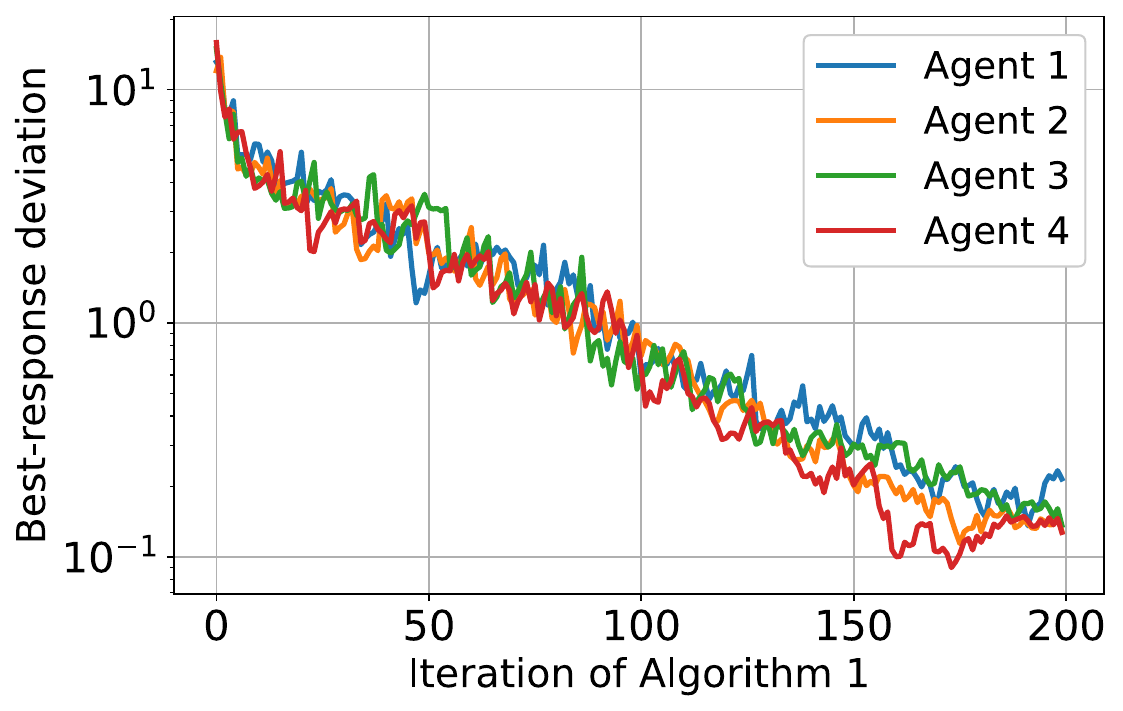}
        \caption{Best response deviation.}
        \label{fig:LQR:BR}
    \end{subfigure}%
    \hfill%%
    \begin{subfigure}[ht]{0.32\textwidth}
        \includegraphics[width=\linewidth]{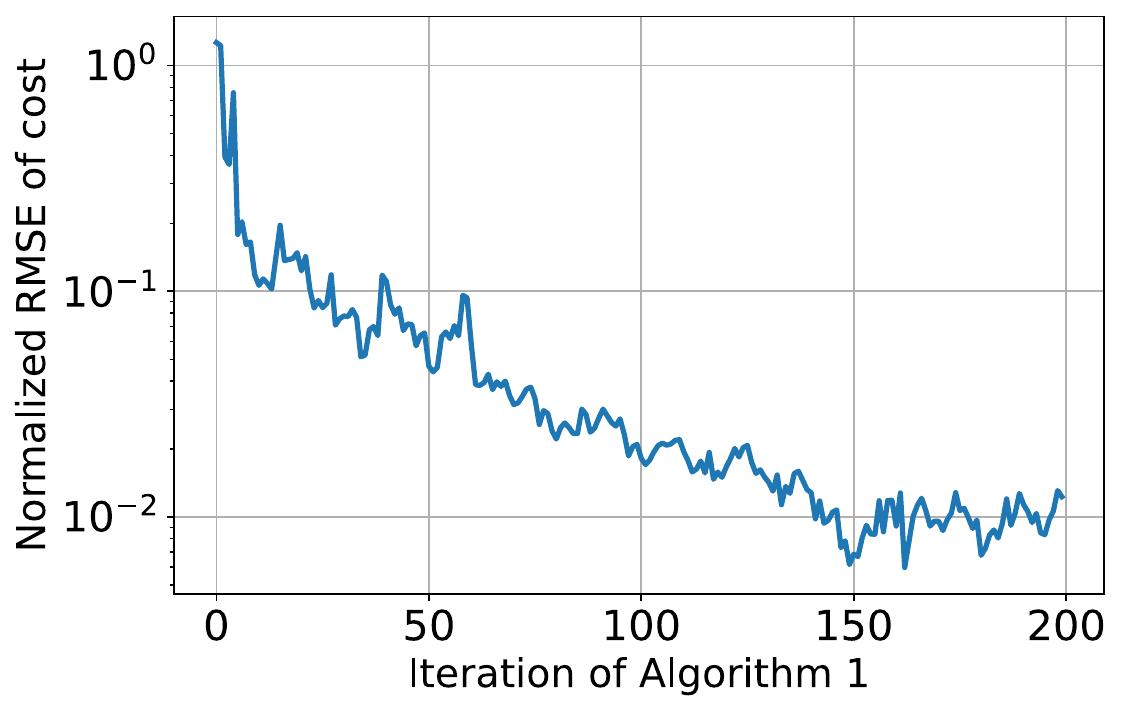}
        \caption{Normalized RMSE of closed-loop cost.}
        \label{fig:LQR:RMSE}
    \end{subfigure}%
    \hfill%%
    \begin{subfigure}[ht]{0.32\textwidth}
        \includegraphics[width=1\linewidth]{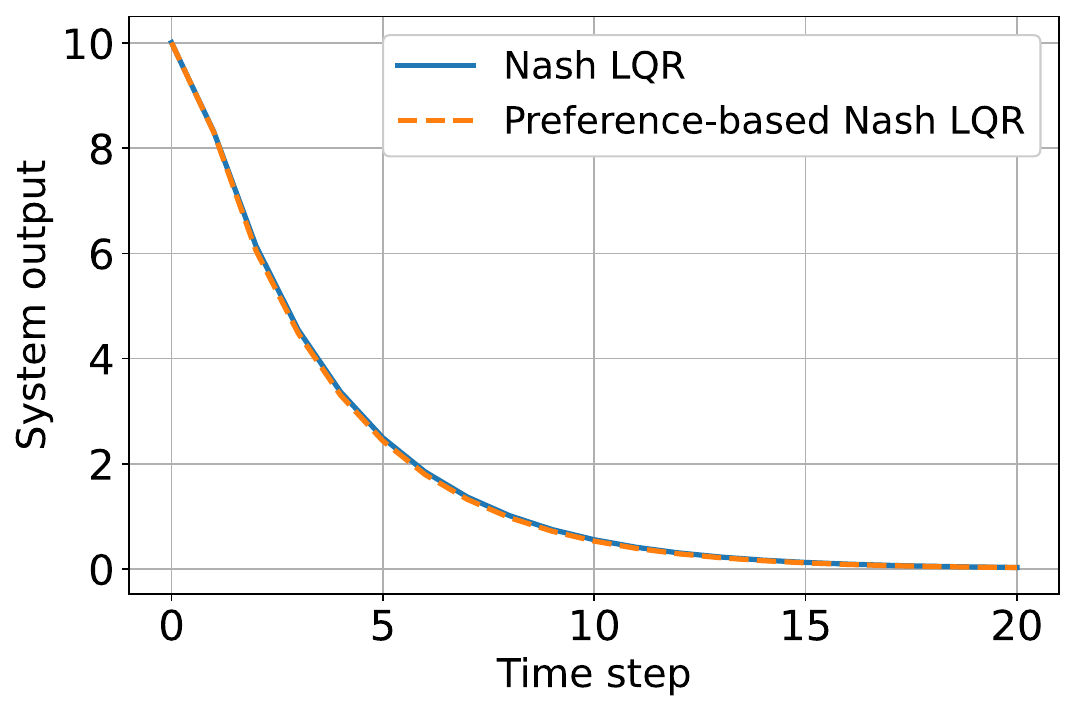}
        \caption{Closed-loop trajectories.}
        \label{fig:LQR:cl}
    \end{subfigure}%
    \hfill%%
    \caption{Results of Algorithm~\ref{alg:AL} for game-theoretic LQR problem with $n_\xi = m = 12$ and $N = 4$.}
    \label{fig:LQR}
\end{figure*}

Table~\ref{tab:LQR} shows the results obtained for the three systems when taking $k_{\rm max} = 100$ and $k_{\rm max} = 200$.
To compare a learned $\hat{K}^{k_{\rm max}}_i$ with the Nash solution $K_i^*$, we take $100$ random initial states.
For each initial state, we perform a closed-loop simulation of length $T$ using $\hat{K}^{k_{\rm max}}_i$ and $K_i^*$, and store the costs obtained with each, as measured using the LQR cost in~\eqref{eq:GNEP:LQR}.
We can then compare $\hat{K}^{k_{\rm max}}_i$ and $K_i^*$ by computing the normalized Root Mean Square Error (RMSE) between the costs obtained in the $100$ tests, where the RMSE is normalized by dividing by the difference between the maximum and minimum costs obtained with $K_i^*$ (so that RMSE values are comparable between the three different system).
Table~\ref{tab:LQR} also shows the maximum value of costs $J_i(\hat{K}_i^{k_{\rm max}}, \hat{K}_{-i}^{k_{\rm max}})$~\eqref{eq:J:NLQR} between the $N$ agents, which measures the deviation of the learned $\hat{K}_i^{k_{\rm max}}$ to a GNE.

Fig.~\ref{fig:LQR} shows the results obtained with the gain matrices $\hat{K}^k_i$ from each iteration $k$ of Algorithm~\ref{alg:AL}, for the case $k_{\rm max} = 200$ and the system with $n_\xi = 12$, $m = 12$, $N = 4$ (see Table~\ref{tab:LQR}).
Fig.~\ref{fig:LQR:BR} shows the best response deviations $J_i$ of each agent, Fig.~\ref{fig:LQR:RMSE} the normalized RMSE (using the same $100$ initial states used for the results in Table~\ref{tab:LQR}), and Fig.~\ref{fig:LQR:cl} shows a closed-loop simulation starting from a random initial state using the GNE gains $K_i^*$ obtained from \texttt{NashOpt} and the learned gains $\hat{K}_i^{k_{\rm max}}$.
We plot the output $y(t)$ of the system, which we take as the summation of the states.
The results shown in Figs.~\ref{fig:LQR:BR} and~\ref{fig:LQR:RMSE} show that the iterates of Algorithm~\ref{alg:AL} converge non-monotonically towards a GNE of the underlying GNEP.
We note that the results obtained for the other examples listed in Table~\ref{tab:LQR} are comparable to the ones shown in Fig.~\ref{fig:LQR}.

\subsection{Solving GNEPs taken from the literature} \label{sec:results:literature}

We present numerical results using Algorithm~\ref{alg:AL} to solve two GNEPs taken from~\cite[Expl. A.3]{Facchinei_report_2009} and~\cite[Expl. 1]{Salehisadaghiani_CoRR_2017}.
For the GNEP from~\cite[Expl. 1]{Salehisadaghiani_CoRR_2017}, we take the configuration presented in~\cite[\S VI.B]{Fabiani_TAC_2024}, which considers $N = 10$ agents.

\begin{figure*}[t]
    \centering
    \begin{subfigure}[ht]{0.23\textwidth}
        \includegraphics[width=\linewidth]{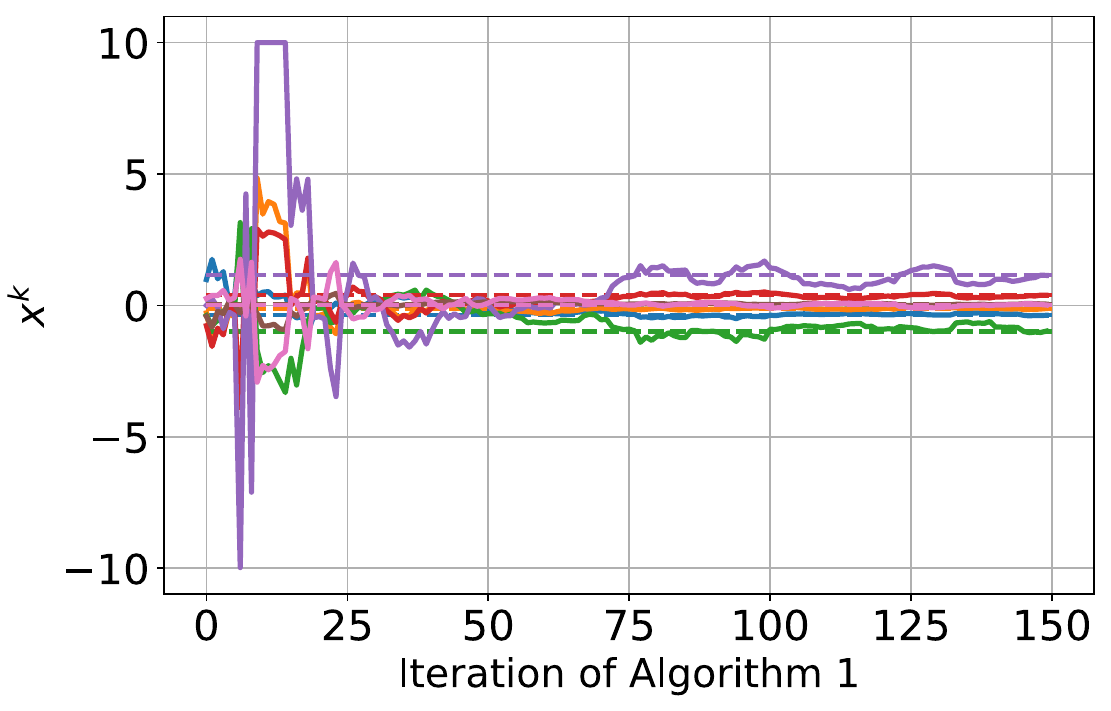}
        \caption{GNEP from~\cite[Expl. A.3]{Facchinei_report_2009}.}
        \label{fig:GNEP:Facchinei_A3:x_pref}
    \end{subfigure}%
    \hfill%%
    \begin{subfigure}[ht]{0.23\textwidth}
        \includegraphics[width=\linewidth]{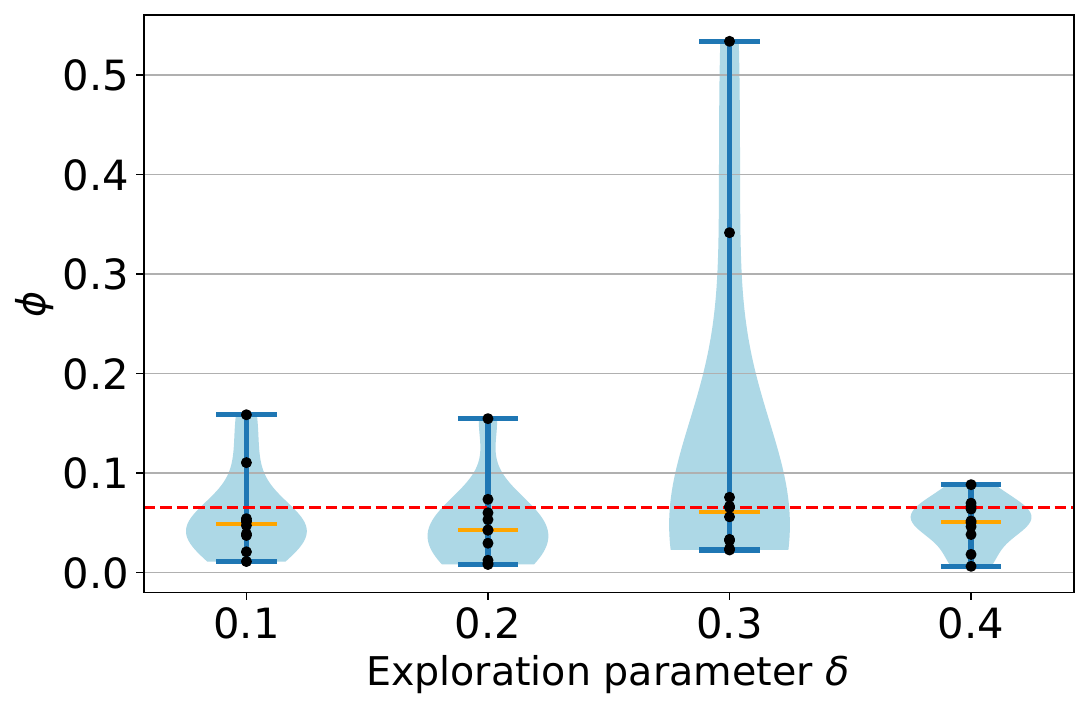}
        \caption{$\phi$ for~\cite[Expl. A.3]{Facchinei_report_2009}.}
        \label{fig:GNEP:Facchinei_A3:br_dev}
    \end{subfigure}%
    \hfill%%
    \begin{subfigure}[ht]{0.23\textwidth}
        \includegraphics[width=1\linewidth]{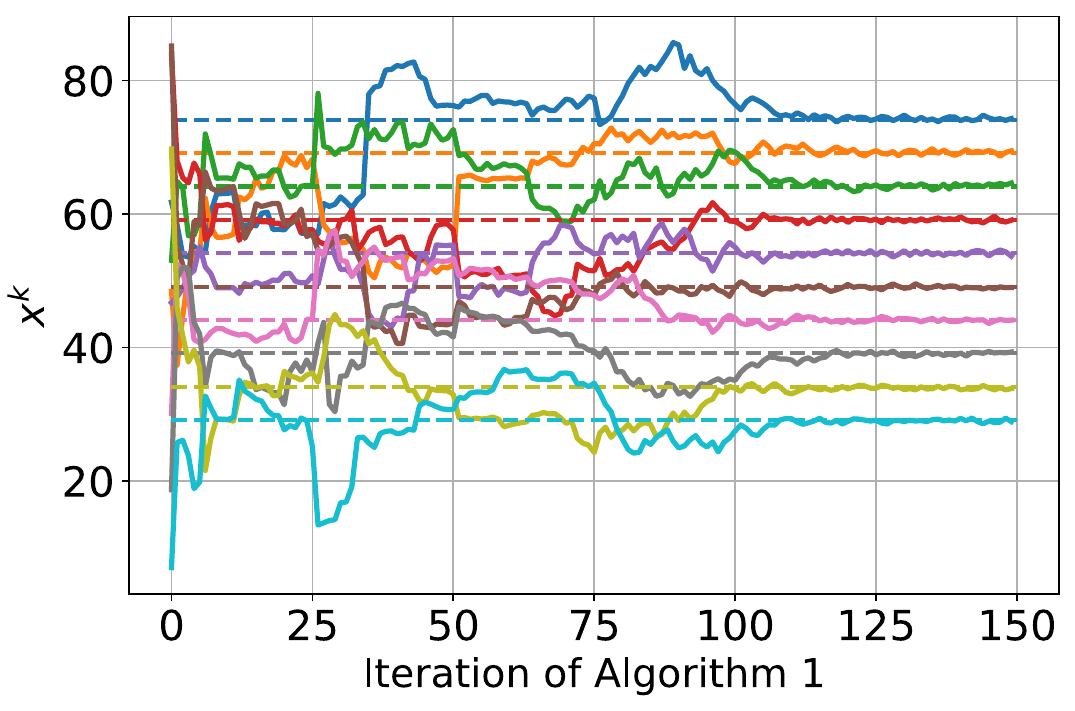}
        \caption{GNEP from~\cite[Expl. 1]{Salehisadaghiani_CoRR_2017} for $N=10$.}
        \label{fig:GNEP:Pavel_Ex1:x_pref}
    \end{subfigure}%
    \hfill%%
    \begin{subfigure}[ht]{0.25\textwidth}
        \includegraphics[width=1\linewidth]{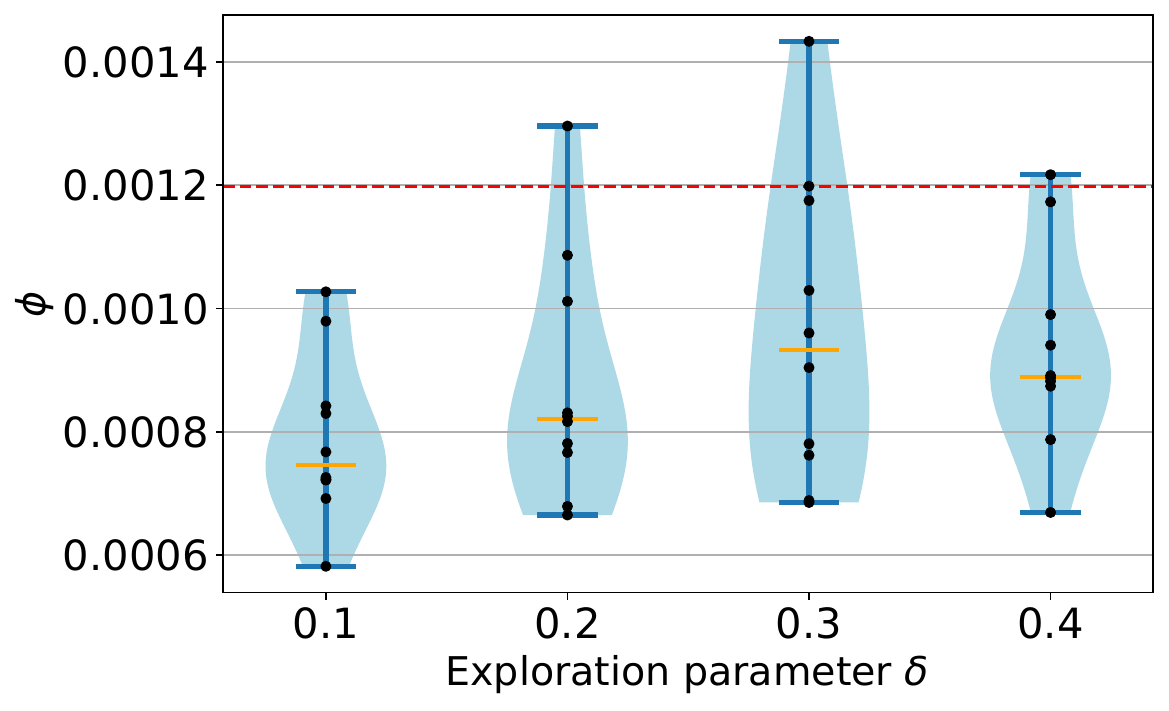}
        \caption{$\phi$ for~\cite[Expl. 1]{Salehisadaghiani_CoRR_2017} for $N=10$.}
        \label{fig:GNEP:Pavel_Ex1:br_dev}
    \end{subfigure}%
    \hfill%%
    \caption{Figures (a) and (c) show the values of $x^k$ obtained from solving~\eqref{eq:GNEPi} for the $\theta_i^k$ learned at each iteration $k$ of Algorithm~\ref{alg:AL} (solid lines) converging towards a GNE of the underlying GNEP (dashed lines). Figures (b) and (d) show the maximum normalized best-response deviations $\phi$ \eqref{eq:max_br_dev} of the tests for different values of $\delta$. Orange lines indicate the median values. For reference, the dashed red lines indicate the value of $\phi$ obtained in Figures (a) and (c).}
    \label{fig:GNEP}
\end{figure*}

Figs.~\ref{fig:GNEP:Facchinei_A3:x_pref} and~\ref{fig:GNEP:Pavel_Ex1:x_pref} show the evolution of the solutions~$x^k$ of problem~\eqref{eq:GNEPi} for the $\theta_i^k$ obtained at each iteration of Algorithm~\ref{alg:AL}, taking $\delta = 0.3$.
The results highlight how the algorithm transitions from an initial phase where the exploration term $\delta^k$ is dominant, towards one where exploitation and local exploration drive $x^k$ towards a GNE of the underlying~GNEP~\eqref{eq:GNEP}.

To evaluate the sensitivity of Algorithm~\ref{alg:AL} to the choice of $\delta$, we test $\delta \in \{0.1, 0.2, 0.3, 0.4\}$, performing $10$ random tests for each value of $\delta$, i.e., with different realization of the initial datasets $\cc{D}_i^0$, exploration points $\bar{x}_i^k$ in~\eqref{eq:simple_exploration}, and noise terms $w_k$ in~\eqref{eq:sigma:noise}.
As seen in Figs.~\ref{fig:GNEP:Facchinei_A3:x_pref} and~\ref{fig:GNEP:Pavel_Ex1:x_pref}, the final iterates of Algorithm~\ref{alg:AL} are influenced by the noise induced by the random variables $\bar{x}_i^k$ and $w^k$.
Therefore, to reduce the effect of the noise, we take the approximate GNE of~\eqref{eq:GNEP} provided by Algorithm~\ref{alg:AL} as the average of the solutions of~\eqref{eq:GNEPi} for the last $5$ iterates $\theta_i^k$ of the algorithm.
Let $\hat{x}^*$ denote the output of Algorithm~\ref{alg:AL} obtained this way and $\tilde{x}_i^*$ its corresponding best response for agent $i$ for the underlying GNEP~\eqref{eq:GNEP}.
For each test, we evaluate the maximum normalized best-response deviation among the $N$ agents:
\begin{equation} \label{eq:max_br_dev}
    \phi(\hat{x}^*) \doteq \max_{i \in \Ni{1}{N}} \frac{\| (\tilde{x}_i^*, \tilde{x}_{-i}^*) - (\hat{x}^*_i, \hat{x}^*_{-i}) \|_2}{\| (\tilde{x}_i^*, \tilde{x}_{-i}^*) \|_2}.
\end{equation}
A value of $\phi(\hat{x}^*) = 0$ indicates that $\hat{x}^*$ is a GNE of~\eqref{eq:GNEP}, with small values of $\phi(\hat{x}^*)$ indicating that $\hat{x}^*$ is a good approximation of a GNE of~\eqref{eq:GNEP}.
Figs.~\ref{fig:GNEP:Facchinei_A3:br_dev} and~\ref{fig:GNEP:Pavel_Ex1:br_dev} present the results of the sensitivity analysis of $\delta$, showing that the proposed method reliably finds good GNE approximations for different values of $\delta$.

\section{Conclusions} \label{sec:conclusions}

We presented an AL method to learn a GNE from preference data only, i.e., with no access to the objective function values or the best responses of the agents. Preferences are used to train the objective functions of a \emph{surrogate} GNEP, whose equilibrium approximates a GNE of the underlying GNEP as the exploitation term of the AL method becomes dominant.
By selecting query points that are close to the current GNE estimate, we train the surrogate functions to form a better local preference classifier.
Numerical results show that, even though the proposed approach is not guaranteed to converge towards a GNE of~\eqref{eq:GNEP}, it is effective at finding good GNE approximations in practice.
An open source implementation of the proposed method can be found at \url{https://github.com/pablokrupa/prefGNEP}.

% Fakesection bibliography
\bibliographystyle{IEEEtran}
\bibliography{IEEEabrv,pref-GNE_bib}

\end{document}